\documentclass[twocolumn,showpacs,preprintnumbers,amsmath,amssymb]{revtex4}
\usepackage{graphicx}
\usepackage{bm}

\def\rd{{\partial}}
\def\sg{{\sigma}}
\def\eps{{\varepsilon}}
\newcommand{\sxy}{\sigma_{\rm SH}}
\newcommand{\oxy}{\sigma_{\rm OH}}
\def\bk{{\bm k}}
\def\ls{{\langle {\bm l} \cdot {\bm s} \rangle_{\mu}}}

\def\w{{\omega}}

\def\g{{\gamma}}
\def\G{{\Gamma}}

\def\th{{\theta}}
\newcommand{\eq}{eqnarray}
\newcommand{\f}{\frac}
\def\nn{{\nonumber}}
\def\V2{{|V_{\bk M \sg}|^2}}
\def\Vk2{{|V_{\bk}|^2}}
\newcommand{\lk}{\left(}
\newcommand{\rk}{\right)}
\newcommand{\tht}{\theta}
\newcommand{\Vs}{V_{\bk M \sg}}
\newcommand{\br}{{\bm{r}}}

\begin{document}

\def\runtitle{Intrinsic Spin and Orbital Hall Effects 
in Heavy Fermion Systems}

\def\runauthor{ T. {\sc Tanaka}$^1$ and H. {\sc Kontani}$^1$
}

%\draft
\title{
Intrinsic Spin and Orbital Hall Effects 
in Heavy Fermion Systems
%based on Periodic Anderson Model
}
\author{ T. {\sc Tanaka} and H. {\sc Kontani}
}

\address{
Department of Physics, Nagoya University,
Furo-cho, Nagoya 464-8602, Japan.
}

\date{\today}

\begin{abstract}
We study the intrinsic spin Hall effect (SHE)
based on the orbitally degenerate periodic Anderson model,
which is an effective model for heavy fermion systems. 
In the very low resistivity regime, the magnitude of the intrinsic 
spin Hall conductivity (SHC) is estimated as 
$2000 \sim 3000 \hbar {\rm e}^{-1} \Omega^{-1} {\rm cm}^{-1}$; 
It is about 10 times larger than that in Pt.
Its sign is negative (positive) in Ce (Yb) compound systems
with $f^1$ ($f^{13}$) configuration.
Interestingly, the obtained expression for the SHC depends only 
on the density of conduction electrons, but is independent of the 
strength of the $c$-$f$ mixing potential and the mass-enhancement factor. 
The origin of the huge SHE is the spin-dependent Berry phase
induced by the complex $f$-orbital wavefunction,
which we call the ``orbital Aharonov-Bohm effect''.
\end{abstract}


\sloppy

\maketitle

%%%%%%%%%%%%%%%%%%
% Introduction
%%%%%%%%%%%%%%%%%%
\section{\label{sec1} Introduction}
%experiment: transition metals including Pt

Spin Hall effect (SHE) is a phenomenon that an applied electric field induces a 
spin current in a transverse direction. 
It has been attracting a great deal of 
interest as a method for creating and detecting spin current. 
Recently, the SHE in metallic systems
%and graphene \cite{Onari}
are intensively studied
due to the interest for both the unsolved origin 
and the possibility of an application to spintronics device
 \cite{Murakami-SHE,Sinova-SHE,Inoue-SHE,Rashba,Raimondi,Saitoh,Valenzuela,ZnSe,Kimura}

Recent intensive studies of the SHE in transition metals was initiated by the 
observation of the huge SHC in Pt \cite{Saitoh,Kimura}.
To elucidate the origin of the huge SHE in transition metals,
theoretical calculations of intrinsic SHE
have been performed intensively 
\cite{Kontani-Ru, Kontani-Pt,Guo-Pt,Tanaka-4d5d}.
The intrinsic SHE occurs in multiband metals with strong 
spin-orbit interaction (SOI) independently of impurities,
which has a close relation to the
intrinsic anomalous Hall effect (AHE) in ferromagnetic metals
\cite{karplus}.
In ref. \cite{Tanaka-4d5d}, the authors have revealed that huge SHEs are ubiquitous in multiorbital
$d$-electron systems by calculating SHEs in various 4$d$ and 5$d$ transition metals.
This study succeeds in explaining sophisticated and systematic
experimental studies by Otani's group \cite{Kimura}.
Therefore,  
it is strongly suggested that the intrinsic mechanism is dominant in transition metals.

The large SHE in transition metals is induced
by the phase factor of the $d$-orbital wavefunction in the presence of
the atomic SOI, which 
we call the ``orbital Aharonov-Bohm (AB) effect'' \cite{Kontani-OHE}.
The intrinsic SHC is predicted to be simply proportional to the 
spin-orbit polarization at the Fermi level
$\langle {\bm l} \cdot {\bm s} \rangle_{\mu}$.
According to the Hund's rule, the SHC should be positive (negative) 
in transition metals with more (less) than half-filling.
Moreover, occurrence of large orbital Hall effect (OHE), 
which is a phenomenon that large $d$-orbital Hall current 
is induced by the electric field,
is also predicted theoretically in many transition metals \cite{Kontani-OHE}.
These fact suggests that a very large 
SHE and OHE may appear in $f$-electron systems
compared to that in $d$-electron systems,
since SHE and OHE are proportional to $\ls$ and $l$, respectively

%AHE is similar phenomena 
In heavy fermion systems, very large AHE appears under the 
magnetic field \cite{Namiki, Otop, Sullow, Hiraoka}:
In clean heavy fermion systems, anomalous Hall conductivity (AHC) $\sxy^a$ is independent of
$\rho$ sufficiently below the coherent temperature $T_0$, whereas $\sxy^a \propto \rho^{-2}$
above $T_0$, which indicates that the intrinsic contribution is dominant in such clean samples.
In ref. \cite{Kontani94}, they studied the AHE based on the orbitally degenerate periodic 
Anderson model (OD-PAM), which is an effective model for heavy fermion compounds.
The obtained general expression has succeeded in explaining the huge AHC observed in 
heavy-fermion systems.
Considering the close relationship between SHE and AHE,
one might expect that huge SHE can be realized in heavy fermion systems.

%Purpose of this paper

In this paper, we study the intrinsic SHE based on the OD-PAM.
It is found that the huge SHE in heavy fermion 
systems originates from the ``orbital AB effect'',
which is given by the spin-dependent Berry phase 
induced by the complex $f$-orbital wavefunction.
In the low resistive regime, the SHCs in Ce- and Yb-compound systems are 
predicted to be about $2000 \sim 3000 
\hbar e^{-1}\cdot \Omega^{-1} \text{cm}^{-1}$ in magnitude, 
which are one order larger than that the value observed in Pt. 
The sign of the SHC is negative (positive) in Ce (Yb) compound systems
with $f^1$ ($f^{13}$) configuration, 
since the SHC is proportional to the spin-orbit polarization 
$\langle {\bm l} \cdot {\bm s} \rangle_{\mu}$ \cite{Kontani-OHE}.
The obtained expression for the SHC does not depend on the strength of 
the $c$-$f$ mixing potential nor the mass-enhancement factor. 
The SHC in $f$-electron systems will be measurable 
by using recently developed fabrication technique 
of high quality heavy fermion thin film \cite{Shishido}.

Recently, present authors have studied the extrinsic SHE based on the orbitally 
degenerate single-impurity Anderson model (OD-SIAM) \cite{Tanaka-NJP}.
Using the Green functional method, we have derived both the skew scattering and side-jump terms analytically.
It is found that the side-jump term derived in the OD-SIAM has a great 
similarity to the intrinsic term derived in the OD-PAM:
The SHCs are simply proportional to $\ls$ 
and their magnitude are almost the same in both mechanisms.
In section \ref{sec:4}, we discuss the relationship between the intrinsic and the side-jump mechanisms.

%%%%%%%%%%%%%%%%%%
% Hamiltonian
%%%%%%%%%%%%%%%%%%
\section{\label{sec:2} Model and Hamiltonian}

In the present paper, we study the intrinsic SHE and OHE
for both  Ce- and Yb-compound heavy fermion systems
based on the OD-PAM.
In these systems, the number of $f$-electron or hole is unity, and the total angular
momentum $J$ is $5/2$ or $7/2$.
In the presence of the strong atomic SOI, the $J=7/2$ level is about 3000 K higher
than the $J=5/2$ level. Therefore, we consider only $J=5/2$ ($J=7/2$) state 
in Ce$^{3+}$ (Yb$^{3+}$) ion with 4$f^{1}$ (4$f^{13}$) configuration.
We note that $\bm{l}\cdot\bm{s}= \f{1}{2} \left[ J(J+1) -L(L+1)-S(S+1) \right]$ is given as follows:
\begin{\eq}
\bm{l}\cdot\bm{s}&=&-2 \ \  \text{for} \ \ J=5/2,  \nn \\ 
\bm{l}\cdot\bm{s}&=&\frac{3}{2} \ \  \text{for} \ \ J=7/2.
\end{\eq}

Here, we introduce the following OD-PAM Hamiltonian,
which had been used to explain the large
Van-Vleck magnetic susceptibility \cite{Kontani-VV} and the 
small Kadowaki-Woods ratio \cite{Kontani-GKWR}
in heavy fermion systems with orbital degeneracy.
\begin{\eq}
\hat H&=&\sum_{\bk\sg} \eps_{\bk}c^{\dagger}_{\bk\sg} c_{\bk\sg} + \sum_{\bk M} E^f f^{\dagger}_{\bk M} f_{\bk M'} \nn \\ 
&&+ \sum_{M \bk \sg} \left( V^{\ast}_{\bk M\sg} f^{\dagger}_{\bk M} c_{\bk \sg}+ V_{\bk M \sg} c^{\dagger}_{\bk \sg}f_{\bk M} \right) \nn \\
&&+U \sum_{i,M\neq M'} n^f_{iM}n^f_{iM'},  \label{eq:Ham}
\end{\eq}
where, $c^{\dagger}_{\bk \sg}$ is the creation operator of a conduction electron 
with spin $\sg=\pm 1$.
$f^{\dagger}_{\bk M}$ is the operator of a $f$-electron with total angular momentum 
$J=5/2 \ (7/2)$ and $z$-component $M \ (-J \leq M \leq J)$ for Ce$^{3+}$ (Yb$^{3+}$). 
$\eps_{\bk}$ is the energy for $c$-electrons,
$E^f$ is the localized $f$-level energy, and $U$ is the Coulomb interaction for $f$-electrons. 
%In the present study, $f$-electron levels are completely degenerate since the 
%magnetic field is absent: $E^f_{M}=E^f$ for all M. 
$V_{\bk M \sg}$ is the mixing potential between the $c$- and $f$-electrons, which
is given by \cite{Kontani94}
\begin{\eq}
V_{\bk M \sg} &=& \sqrt{\f{2}{2J+1}} \sqrt{4\pi} V_f  \sum_{m} a^M_{m\sg} 
Y^m_{l}(\th_{\bk}, \phi_{\bk}), 
\end{\eq}
where, $a^M_{m \sg}$ is the Clebsh-Gordan (C-G) coefficient and $Y^m_l(\th_{\bk}, 
\phi_{\bk})$ is the spherical harmonic function. 
Here, the C-G coefficient for $l=3$ is given by (for l=3)
\begin{\eq}
a^M_{m\sg} &=& -\sg \left\{ \left( 7/2-M\sg \right)/7  \right\}^{1/2} \delta_{m,M-\sg/2} \ \  \text{for} \ \ J=5/2,  \nn \\ 
a^M_{m\sg} &=& \left\{ \left( 7/2+M\sg \right)/7  \right\}^{1/2} \delta_{m,M-\sg/2} \  \qquad \text{for} \ \ J=7/2. \label{eq:J7/2} \nn \\
\end{\eq}
Here, the $\bk$-dependence of $V_f$ is neglected due to the small radius of the $f$-orbital wave function.
%($A$ is the $T^2$-coefficient of the resistivity,
%and $\gamma$ is the electric specific heat coefficient.)
We also neglect the crystalline electric field splitting of $E^f$-level
since its effect on the intrinsic Hall effect would not be essential
 \cite{AHE-CEF}.
Hereafter, we put $U=0$; the effect of Coulomb interaction on the SHC will be discussed in section \ref{sec:4}.

From the expression of the C-G coefficient in eq. (\ref{eq:J7/2}), we see that  
conduction electrons with $\uparrow$-spin mainly hybridize 
%strongly couples 
with $M=-5/2$ ($M=7/2$) for $J=5/2$($J=7/2$), 
which is consistent with the Hund's rule: That is the spin and orbital angular momentum
are parallel (antiparallel) for $J=5/2$ ($J=7/2$). 
We will show that the sign of the SHC is explained by the spin-orbit polarization \cite{Kontani-OHE}. 
In the present study, we neglect the effect of crystalline electric field on $f$-orbitals, since it
is small due to the small radius of the $f$-orbital wave function. Hereafter, 
we put $\hbar =1 $.

\begin{figure}[!htb]
\includegraphics[width=.5\linewidth]{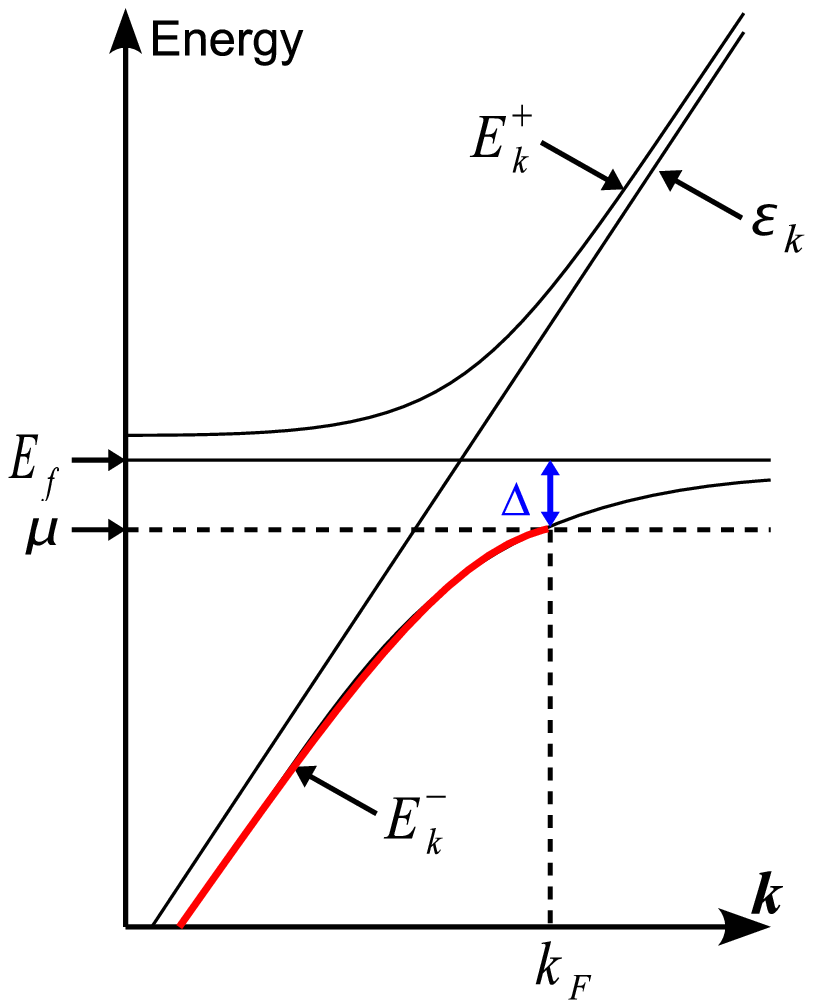}
\caption{\label{fig:band} Band structure of the OD-PAM given in eq. (\ref{eq:Ham}).
Here, $E^{\pm}_{\bk}$ is the hybridization band. 
} 
\end{figure}

In Fig. \ref{fig:band}, we show the band structure of OD-PAM given in eq. (\ref{eq:Ham}).
Here, $E^{\pm}_{\bk}$ represents the hybridization bands given by 
$E^{\pm}_{\bk}=\f{1}{2} \left[ (\eps_{\bk} + E^f) \pm 
\sqrt{(\eps_{\bk}-E^f)^2+ 4|V_f|^2}  \right]$. 
%Note that we used the following 
%relationship: $\sum_{M} |V_{\bk M \sg}|^2 = 3|V_{\bk}|^2$.
In this study, we assume the 
metallic state, where the Fermi level $\mu$ lies in the $c$-$f$ hybridization band.
In this figure, $k_{F}$ is the Fermi momentum and $\Delta\equiv E^f - \mu$. 

%%%%%%%%%%%%%%%%%%%%%%%%%%%%%%%%%%%%%%%%%%
%%%%%%%%%%%%%%%%%%%%%%%%%%%%%%%%%%%%%%%%%%

%%%%%%%%%%%%%%%%%%
% Green function
%%%%%%%%%%%%%%%%%%

%%%%%%%%%%%%%%%%%%%%%%%%%%%%%%%%%%%%%%%%%%
\begin{figure}[!htb]
\includegraphics[width=.9\linewidth]{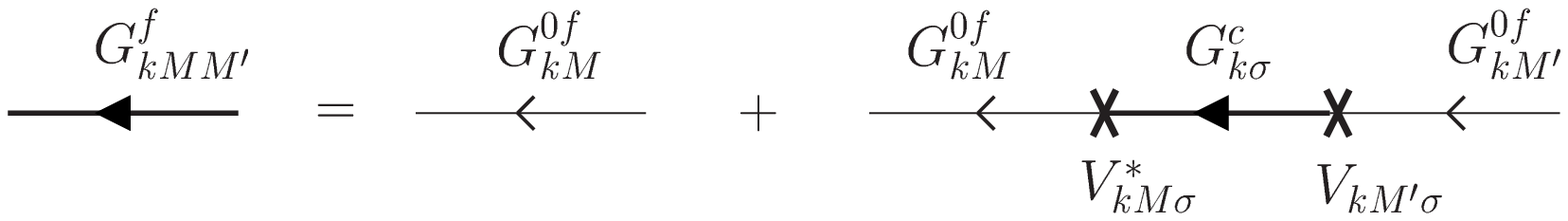}
\caption{\label{fig:Gf} The diagrammatic expression for the Green function 
in eq. (\ref{eq:Gf}) \cite{Kontani94}.
} 
\end{figure}
%%%%%%%%%%%%%%%%%%%%%%%%%%%%%%%%%%%%%%%%%%
Here, the conduction and $f$-electron Green functions for OD-PAM 
in the absence of the magnetic field are given by as follows \cite{Kontani94}:
\begin{\eq}
G^c_{\bk \sg \sg}(\w) &=&\left( \w+\mu -\eps_{\bk} - \sum_{M} \f{\V2}{\w+\mu-E^f_{M}} \right)^{-1}, \\
G^f_{\bk M M'}(\w) &=& G^{0f}_{\bk M}(\w) \delta_{MM'} \nn \\
&+& \sum_{\sg} G^{0f}_{\bk M}(\w) V^{\ast}_{\bk M \sg} G^c_{\bk \sg}(\w) V_{\bk M'\sg} G^{0f}_{\bk M'}(\w).  \nn \\ \label{eq:Gf}
\end{\eq}
We note that $G^c_{\bk \sg \bar \sg}(\w)=0$ \cite{Kontani94}.  
The diagrammatic expression for eq. (\ref{eq:Gf}) is given in Fig. \ref{fig:Gf}.
$G^{0f}$is the $f$-electron Green function without hybridization given as
\begin{\eq}
G^{0f}_{\bk M}(\w)&=& \f{1}{\w+\mu-E^f}.
\end{\eq}

Now, we consider the quasiparticle damping rate $\hat \Gamma(\w)$, which is mainly given by the imaginary part of the $f$-electron self-energy, 
$\hat \Sigma_{\bk}(\w)$ in heavy fermion systems.
In the dynamical mean-field approximation (DMFA), the 
self-energy is composed local $f$-Green function, $\f{1}{N} \sum_{\bk} G_{\bk M M'}(\w)\equiv g(\w)\delta_{MM'}$, which is diagonal with respect to $M$ and is dependent of $M$ in the orbitally degenerate case \cite{Kontani-GKWR}.
Here, $N$ is the number of $\bk$-points.
Therefore, in the present study, we assume that $\hat \Gamma$ is diagonal with respect to $M$, and is independent of the momentum. Moreover, since $f$-electrons
are degenerate in the present model, we assume that $\Gamma_{M}$ is 
approximately independent of $M$ and can be approximated as
$\Gamma_{MM'}=\gamma\delta_{MM'}$, where $\g$ is a constant. In this study, we perform a calculation of the SHC
using this constant $\g$ approximation. Then, the retarded (advanced) Green functions are given by
\begin{\eq}
G^{c\rm{R(A)}}_{\bk}(\w)&=&\left( \w+\mu-\eps_{\bk} -\f{|V_f|^2}{\w+\mu-E^f +(-) i\g} \right)^{-1}, \nn \\
G^{0f\rm{R(A)}}_{\bk}(\w)&=&\left( \w+\mu -E^f +(-)i \g \right)^{-1}.
\end{\eq}

%%%%%%%%%%%%%%%%%%
% Kubo formula
%%%%%%%%%%%%%%%%%%
\section{\label{sec:3} Calculations of SHC and OHC}

In this study, we calculate $\sxy$ based on 
linear response theory. 
According to Streda \cite{Streda}, the SHC at $T=0$ 
in the absence of the current vertex correction (CVC)
is given by $\sxy = \sxy^{I} + \sxy^{II}$, 
where 
\begin{\eq}
\sxy^{I} &=& \f{1}{2\pi N} \sum_{\bk} \text{Tr} \left[ \hat J^{\rm{S}}_x \hat G^{\rm{R}} \hat J^{\rm{C}}_y \hat G^{\rm{A}} \right]_{\w=0}, \label{eq:FS-term} \\
\sxy^{II} &=& \f{-1}{4\pi N} \sum_{\bk} \int_{-\infty}^{0} d\w \text{Tr} \left[ \hat J^{\rm{S}}_x \f{\rd \hat G^{\rm{R}}}{\rd \w} \hat J^{\rm{C}}_y \hat G^{\rm{R}} \right. \nn \\
&& \qquad \left. - \hat J^{\rm{S}}_x \hat G^{\rm{R}} \hat J^{\rm{C}}_y \f{\rd \hat G^{\rm{R}}}{\rd \w} -\langle \rm{R}\leftrightarrow \rm{A} \rangle \right]. \label{eq:sea-term}
\end{\eq}
%where $N$ is the number of $\bk$-points.
Here,
$\sxy^{I}$ and $\sxy^{II}$ represents the Fermi surface term and the Fermi sea term, respectively. 

%%%%%%%%%%%%%%%%%%
% current
%%%%%%%%%%%%%%%%%%
In the present model, the charge current operator is given by $\hat J^C_{ \mu}=-e \hat v_{\bk \mu} $, where $-e \ (e>0)$ is the electron charge, and  
\begin{\eq}
\hat v_{\bk \mu} &=& \sum_{\sg} \f{\rd}{\rd k_{\mu}} \eps_{\bk} c^{\dagger}_{\bk \sg}c_{\bk \sg} 
\sum_{\sg M} \left\{ \f{\rd}{\rd k_{\mu}} V_{\bk M \sg} c^{\dagger}_{\bk \sg}f_{\bk M} + {\rm h.c} \right\}. \label{eq:velocity} \nn \\
%&&+ \f{\rd}{\rd k_{\mu}} V^{\ast}_{\bk M \sg} f^{\dagger}_{\bk M} c_{\bk \sg}.
\end{\eq}
Next, we explain the $s_z$-spin current operator $\hat J^S_{ \mu}$.
In the present model, $\hat s_z$ is given by
\begin{\eq}
\hat s_{z}&=& \sum_{\sg}\f{\sg}{2}c^{\dagger}_{\bk\sg}c_{\bk\sg} + \sum_{M} S_{M}
f^{\dagger}_{M \bk} f_{M \bk},
\end{\eq}
where $S_{M}=\sum_{m \sg} \f{\sg}{2} \left[ a^M_{m\sg} \right]^2 $.
It is straight forward to show that $S_{M}=-\f{M}{7}$ ($\f{M}{7}$) for $J=5/2$ $(J=7/2)$.
Then, the spin current 
$\hat J^S_{ \mu}\equiv\left\{ \hat v^c_{\bk \mu}, \hat s_z \right\}/2$ is given by
\begin{\eq}
\hat J^S_{ \mu} &=& \sum_{\sg}\f{\sg}{2} \f{\rd \eps_{\bk}}{\rd k_{\mu}} c^{\dagger}_{\bk\sg} c_{\bk \sg} \nn \\
&+&  \sum_{\sg M}\left\{\f{1}{2} \left( \f{\sg}{2} + S_M \right) \f{\rd V_{\bk M \sg}}{\rd k_{\mu}} c^{\dagger}_{\bk \sg} f_{\bk M} + \rm{h.c} \right\}. \label{eq:spcurrent} \nn \\
%&&+ \f{1}{2} \left( \f{\sg}{2} + S_M \right) \f{\rd V^{\ast}_{\bk M \sg}}{\rd k_{\mu}} f^{\dagger}_{\bk M} c_{\bk \sg} 
\label{eq:spin-current}
\end{\eq}

In a similar way, the total angular momentum current operator,
$\hat J^J_{ \mu}\equiv\left\{ \hat v^c_{\bk \mu}, \hat J_z \right\}/2$,
is given by replacing $S_M$ in eq. (\ref{eq:spin-current}) with $M$.
Then, the the orbital angular momentum current operator,
$\hat J^O_{ \mu} \equiv \hat J^J_{ \mu}-\hat J^S_{ \mu}$,
is expressed as
\begin{\eq}
\hat J_{\mu}^{O}&=&\left\{ \hat v^c_{\bk \mu}, \hat l_z \right\}/2 \nn \\
&=& \sum_{\sg M}\left\{ \f{1}{2}(M-S_M) \f{\rd V_{\bk M \sg}}{\rd k_{\mu}} c^{\dagger}_{\bk \sg} f_{\bk M}  + \rm{h.c} \right\}.
\end{\eq}
Then, the orbital Hall conductivity (OHC)
$\sigma_{\rm OH}\equiv \langle J_{x}^{O} \rangle/E_y$ due to the OHE 
is given by $\sigma_{\rm OH} = \sigma_{\rm OH}^{I} + \sigma_{\rm OH}^{II}$,
where $\oxy^I $ and $\oxy^{II}$ are respectively given by 
eqs. (\ref{eq:FS-term}) and (\ref{eq:sea-term}) 
by replacing $J^S_x$ with $J^O_x$.
% which is given by$\hat J^J_{\mu}=\left\{ \hat v^c_{\bk \mu}, \hat J_z \right\}/2$.
%In the present model, $\hat J^J_{\mu}$ is given by eq. (\ref{eq:spin-current}) 
%by replacing $S_M$ with $M$. 

\begin{figure}[!htb]
\includegraphics[width=.5\linewidth]{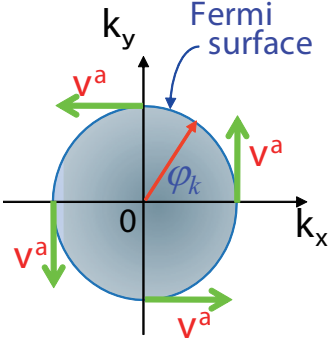}
\caption{\label{fig:anomalous} A schematic view of the anomalous velocity $v^a$.
} 
\end{figure}

Here, we study the velocity given by 
the $c$-$f$ mixing potential $V_{\bk M \sg}$ \cite{Kontani94}:
\begin{\eq}
\f{\rd V_{\bk M \sg}}{\rd k_x} 
&=& -i \left( M-\f{\sg}{2} \right) \f{k_y}{k_x^2 + k_y^2} V_{\bk M \sg} 
 \nonumber \\
& &+ \frac{\rd}{\rd k_x} \left(V_{\bk M \sg}\alpha_{M,\sg}^*\right)\alpha_{M,\sg}
 \nonumber \\
&\equiv& v_x^a + v_x^b
\label{eq:anomalous}.
\end{\eq}
Here, $v_x^a$ is the anomalous velocity given by $\bk$-derivative of 
the phase factor 
$\alpha_{M,\sg}=\rm{exp} \left\{i\left(M-\f{\sg}{2}\right)\phi_k\right\}$ 
in $V_{\bk M \sg}$.
Figure \ref{fig:anomalous} is a schematic view of the anomalous velocity
$v^a \propto \nabla_\bk \phi_{\bk}$.
Since $v^a_x \propto k_y$ and thus 
$\sum_\bk v^a_x (\partial\epsilon_\bk/\partial k_y)\ne0$, 
the anomalous velocity gives rise to the large SHE and AHE 
in heavy fermion systems.
On the other hand, $v^b_x \propto k_x$ gives a normal velocity.
In eqs. (\ref{eq:FS-term}) or (\ref{eq:sea-term}),
the terms which contain single $v_\mu^a$ give rise to the SHC.

%%%%%%%%%%%%%%%%%%%
%%%%%%%%%%%%%%%%%%%%%%%%

%%%%%%%%%%%%%%%%%%
% Calculations of FS term
%%%%%%%%%%%%%%%%%%

\subsection{\label{subsec:3-1} Calculation of the Fermi surface term}
%rrrrrrrrrrrrr
Here, we calculate the SHC by neglecting CVC according to eqs. (\ref{eq:FS-term}) and (\ref{eq:sea-term}), using eqs. (\ref{eq:velocity}) and (\ref{eq:spcurrent}).
$J^C_{\mu}$ and $J^S_{\nu}$ are composed of  the conduction electron term
$\rd \eps_{\bk}/\rd k_{\mu}  \equiv \rd_{\mu} \def \rd_{\mu} \eps_{\bk}$ and the hybridization term $\rd_{\mu} V_{\bk}$.
Fig. \ref{fig:diagram} shows the terms for $\sxy$ in which
$\hat J^S_x, \hat J^C_y$ is composed of zero or one $\rd_{mu} V_{\bk}$.
Fig. \ref{fig:diagram} (a) gives large SHC since $\rd_{\mu} V_{\bk}$ includes the 
anomalous velocity in eq. (\ref{eq:anomalous}).
%the anomalous velocity.
% whose diagrammatic expressions are shown in Fig. \ref{fig:diagram} (a).
We note that the terms in \ref{fig:diagram} (b) that are composed only of $\rd_{x} \eps_{\bk} \cdot \rd_{y} \eps_{\bk}$ vanishes identically.
Moreover, there exists the terms that
are proportional to $\rd_{\mu}V_{\bk}\rd_{\nu}V_{\bk}$,
as shown in Fig\ref{fig:diagram2}.
%
%both current operators are composed of the anomalous velocity, \
In Appendix B, we will show that 
%the contribution of 
these terms are much smaller than the contribution by Fig. \ref{fig:diagram} (a). 
%the terms either $\hat J^S_x$ or $\hat J^C_y$ is composed of the anomalous velocity.
Therefore, we here focus on the terms in Fig. \ref{fig:diagram} (a).

In this subsection, we derive the analytical expression for the Fermi surface term, 
since the Fermi surface term dominates over the Fermi sea term,
as discussed in previous studies 
\cite{Kontani06,Kontani-Ru,Kontani-Pt,Tanaka-4d5d}.
The Fermi sea term will be derived in section \ref{subsec:3-2}.

%%%%%%%%%%%%%%%%%%
% Diagrams
%%%%%%%%%%%%%%%%%%
%%%%%%%%%%%%%%%%%%%%%%%%%%%%%%%%%%%%%%%%%%
\begin{figure}[!htb]
\includegraphics[width=1\linewidth]{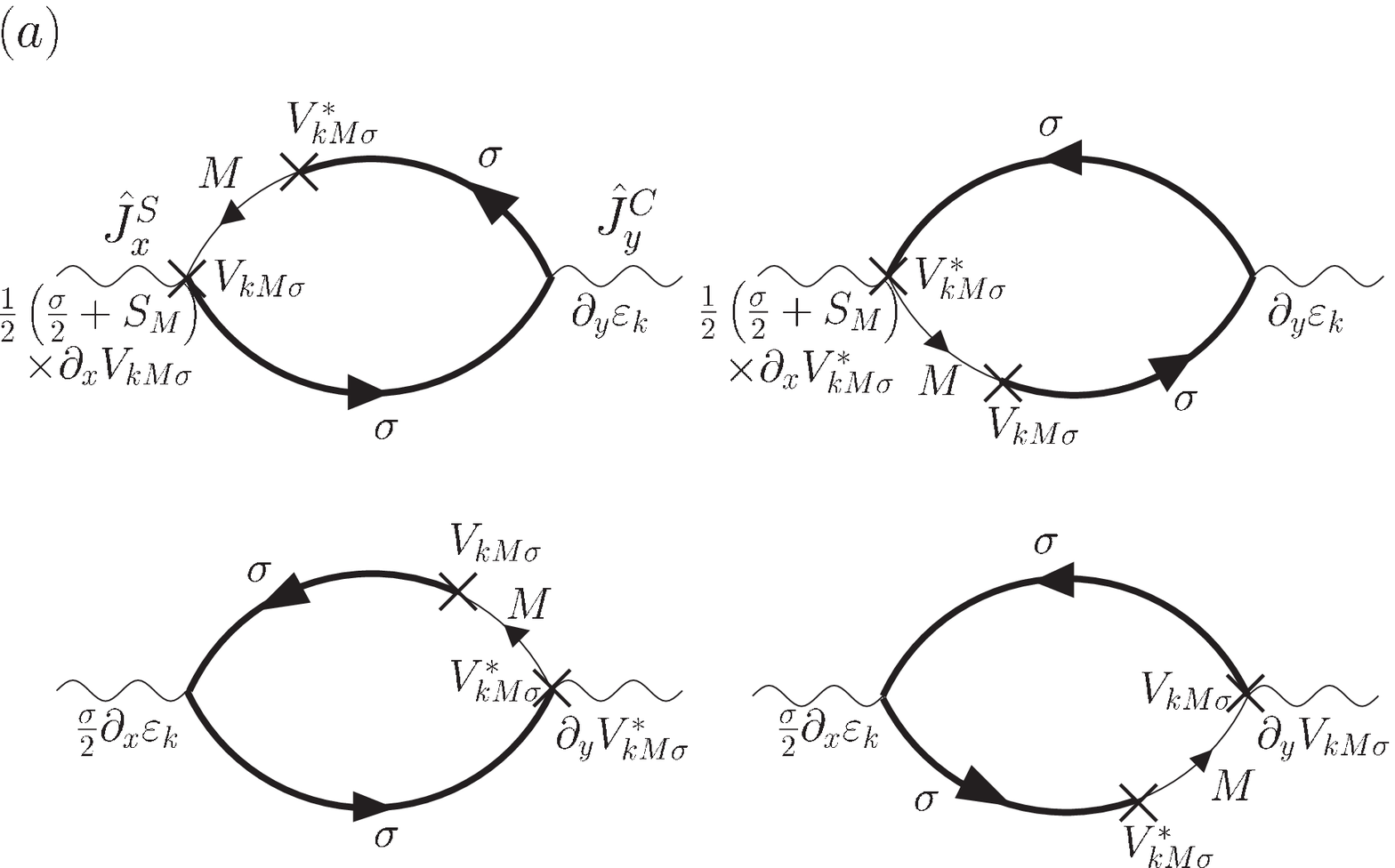} 
\includegraphics[width=.6\linewidth]{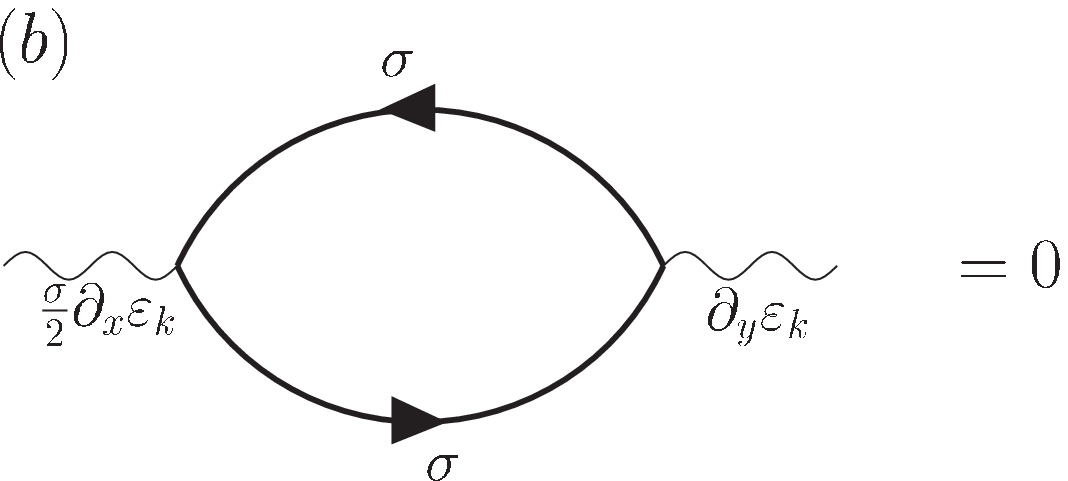} 
\caption{\label{fig:diagram} The diagrammatic expressions for $\sxy$.
(a) The diagrammatic expressions for the dominant terms. (b) The diagrammatic expressions of the terms composed only of $\rd_{\mu} \eps_{\bk}$,
which vanishes identically.
} 
\end{figure}
%%%%%%%%%%%%%%%%%%%%%%%%%%%%%%%%%%%%%%%%%%

According to eqs. (\ref{eq:FS-term}), (\ref{eq:velocity}) and (\ref{eq:spcurrent}), the Fermi surface term $\sxy^{I}$ for
%, which is represented in 
Fig. \ref{fig:diagram} (a) is given by
\begin{\eq}
\sxy^I &=& \f{-e}{2\pi N} \sum_{\bk M \sg} \f{1}{2} \left( \f{3\sg}{2} + S_M \right) \nn \\
&\times&\left[ \f{\rd V_{\bk M \sg}}{\rd k_x} \f{\rd \eps_{\bk}}{\rd k_y} V^{\ast}_{\bk M \sg} |G^{cR}_{\bk}(0)|^2G^{0fR}_{\bk} (0) + {\rm c.c.} \right].
% &+ & \left. \f{1}{2} \left( \f{3\sg}{2} + S_M \right) \f{\rd V^{\ast}_{\bk M \sg}}{\rd %k_x} \f{\rd \eps_{\bk}}{\rd k_y} V_{\bk M \sg} |G^c_{\bk}(0)|^2G^{fA}_{\bk}(0) \right],  
\label{eq:sxyI-1} \nn \\
\end{\eq}
%where $|G^c_{\bk}(0)|^2=G^{cR}_{\bk}(0)G^{cA}_{\bk}(0)$.

Here, we confine ourselves to the case $J=5/2$ state corresponding to Ce
$^{3+}$-ion. In section \ref{subsec:4-1}, we will discuss the case for $J=7/2$ 
state.
Then, by using the following relationships
\begin{\eq}
&&\sum_{M \sg} M^2 \V2 = \f{|V_f|^2}{2} \left( 1 + 16\sin^2\th \right), \label{eq:Ms1} \\
&&\sum_{M \sg} \sg^2 \V2 = 2 |V_f|^2,  \\
&&\sum_{M \sg} M \sg \V2 =  |V_f|^2 \left( 1 - 4\sin^2\th \right), \label{eq:Ms3} \\
&&\f{k_y}{k_x^2 + k_y^2} = \f{1}{k} \f{\sin\th \sin\phi}{\sin^2\th}, \ \f{\rd \eps_{\bk}}{\rd k_y} = \f{\rd \eps_{\bk}}{\rd k} \sin\th \sin\phi,
\end{\eq}
eq. (\ref{eq:sxyI-1}) is transformed as follows:
\begin{\eq}
\sxy = \f{-e}{2\pi N} \f{52}{7} |V_f|^2 \sum_{k} \f{1}{k} \f{\rd \eps_{\bk}}{\rd k} \f{\g}{(\mu-E_{\bk})^2+\g^2} |G^c_{\bk}(0)|^2, \label{eq:sxyI-2} \nn \\
\end{\eq}
where $k\equiv |\bk|$.

Here, we analyze eq. (\ref{eq:sxyI-2}) when $\g$ is small enough: In this case,
\begin{\eq}
G^{cR}_{\bk}(0)&=& \left( \mu-\eps_{\bk} -\f{|V_f|^2}{\mu -E^f + i\g} \right)^{-1} \nn \\
 &\simeq & \left( \mu- \tilde \eps_{\bk} +i\G_c \right)^{-1},
\end{\eq}
where $\displaystyle \tilde \eps_{\bk} = \eps_{\bk} + \f{|V_f|^2}{\mu - E^f}$, and $\displaystyle \G_c= \f{|V_f|^2}{(\mu-E^f)^2}\g$.
Since $\gamma/(x^2+\gamma^2) = \pi \delta(x)$ for small $\gamma$,
we obtain the following relationship:
\begin{\eq}
|G^{cR}_{\bk}(0)|^2 &=& \f{1}{(\mu - E_{\bk})^2 + \G_c^2} \nn \\
&\simeq& \f{\pi}{\G_c} \delta(\mu-E_{\bk}).
\end{\eq}

Substituting above equation into eq. (\ref{eq:sxyI-2}), we obtain the 
following relationship for small $\g$:
\begin{\eq}
\sxy &=& \f{-e}{2\pi N } \f{52}{21} \sum_{k} \f{1}{k} \f{\rd \eps_{\bk}}{\rd k} \delta(\mu-E_{\bk}). \label{eq:sxyI-3}
\end{\eq}
Now, we approximate the conduction electron as free electron.
Then, $\sxy^{I}$ for $J=5/2$ is given by
\begin{\eq}
\sxy^{I}&=& -e\f{26}{21} \f{k_F}{2\pi^2} N_{FS} 
 \nonumber \\
&=& -\f{e}{2\pi a} \f{26}{21} N_{FS}  \label{eq:I-va},
\end{\eq}
where $a$ is the lattice spacing and
$N_{FS}$ represents the number of large Fermi surface. 
The first line in eq. (\ref{eq:I-va}) means that the SHC
depends only on the density of conduction electron
$n_{\rm c}=k_{\rm F}^3/3\pi^2$, except for $N_{FS}$.
This result suggests that SHCs in $Ce$-compound heavy fermion systems
take similar large negative values.
The second line in eq. (\ref{eq:I-va}) is obtained 
by putting $k_{\rm F}=\pi/a$.
When $a=4$\AA, then $e/2\pi a \approx 1000 \hbar e^{-1}\Omega^{-1} \rm{cm} $.
If we assume that $N_{FS}=2\sim 3$, we obtain $\sxy = 2000\sim 3000 \hbar e^{-1} \Omega^{-1}$cm$^{-1}$ for Ce-compound system.
%Here, we note that the sign of $\sxy^{zI}$ is negative for $J=5/2$.
Interestingly, the expression obtained above is independent of the 
strength of the $c$-$f$ mixing potential.
% nor the renormalization factor. 

%%%
%OHC
%%%

Next, we discuss the Fermi surface term for the OHC. 
By replacing $\hat J^S_x$ with $\hat J^O_x$ in eq. (\ref{eq:FS-term}), 
$\sigma^{I}_{\rm OH}$ can be calculated in the same way as SHC.
The obtained result is 
\begin{\eq}
\sigma^{I}_{\rm OH} = -\f{7}{13} \sxy^I.
 \label{eq:O-Ce}
\end{\eq}
Thus, $\sigma^{I}_{\rm OH}$ shows a large positive value in Ce-compounds.
In contrast, the relation $\sigma^{I}_{\rm OH} \gg |\sxy^I|$
is satisfied in transition metals since the SOI is weak and $\ls\ll1$
\cite{Tanaka-4d5d}.

%%%%%%%%%%%%%%%%%%%%%%%
%Calculations of Fermi sea term
%%%%%%%%%%%%%%%%%%%%%%%
\subsection{\label{subsec:3-2} Calculation of the Fermi sea terms}

In this section, we derive the analytical expression for the Fermi sea term $\sxy^{II}$, and show that the Fermi surface term ($I$) dominates the Fermi sea term ($II$).

\begin{widetext}
According to eqs. \ref{eq:sea-term}, \ref{eq:velocity}, and \ref{eq:spcurrent}, the Fermi surface term $\sxy^{II}$ for Fig. \ref{fig:diagram} (a) is given by%, which corresponds to the cV-terms, is given by
\begin{\eq}
\sxy^{II}&=& \f{e}{4\pi N} \sum_{\bk M \sg} \int^{0}_{-\infty} d\w \ \f{1}{2} \left( \f{3\sg}{2} + S_M \right) \nn \\
&\times& \left[ \f{\rd V_{\bk M \sg}}{\rd k_x} \f{\rd\eps_{\bk}}
{\rd k_y} \left\{ \f{\rd G^{fR}_{\bk}(\w)}{\rd \w} G^{cR}_{\bk}(\w) -G^{fR}_{\bk}(\w) \f{\rd G^{cR}_{\bk}(\w)}{\rd \w}  - \langle R\leftrightarrow A \rangle  \right\} \right. \nn \\
&&+\left. \f{\rd V^{\ast}_{\bk M \sg}}{\rd k_x} \f{\rd \eps_{\bk}}{\rd k_y} \left\{ \f{\rd G^{cR}_{\bk}(\w) }{\rd \w} G^{fR}_{\bk}(\w) - G^{cR}_{\bk}(\w) \f{\rd G^{fR}_{\bk}(\w)}{\rd \w} - \langle R\leftrightarrow A \rangle \right\} \right].  \label{eq:sxyII-1}
\end{\eq}
\end{widetext}

Using the relations in eqs. (\ref{eq:Ms1}) - (\ref{eq:Ms3}), and performing the $M,\sigma$-summations in eq. (\ref{eq:sxyII-1}), it is transformed as
\begin{\eq}
\sxy^{II} &=& \f{-e}{4\pi N} \sum_{\bk} \left(-\f{52}{7} \right) |V_f|^2\f{1}{k}
\f{\rd \eps_{\bk}}{\rd k} \nn \\
&&\text{Im}  \left\{ \int_{-\infty}^{0} \f{d\w}{\left[ (\w-\eps_{\bk})(\w-E^f+i\g)-|V_f|^2 \right]^2} \right\}.  \label{eq:sxyII-2} \nn \\
\end{\eq}
To perform the $\w$-integration in eq. (\ref{eq:sxyII-2}), we rewrite the integrand in eq. (\ref{eq:sxyII-2}) as follows:
\begin{\eq}
&&\left( \w+\mu-\eps_{\bk} \right)\left( \w+\mu-E^f+i\g \right) -|V_f|^2 \nn \\
&=& \left(\w+\mu -E^{+}_{\bk}+i\g^+ \right)\left( \w+\mu-E^{-}_{\bk}+i\g^- \right),
\end{\eq}
where 
\begin{\eq}
\g^{\pm}= \f{\g}{2} \left( 1\mp \f{\eps_{\bk}-E^f}{E^+_{\bk} - E^-_{\bk}} \right).
\end{\eq}
Then, the $\w$-integration in eq. (\ref{eq:sxyII-2}) can be performed analytically as follows:
\begin{widetext}
\begin{\eq}
&&\int_{-\infty}^{0} \f{d\w}{\lk \w+\mu-E^+_{\bk} +i\g^+ \rk^2 \lk \w+\mu -E^-_{\bk} +i\g^- \rk^2 } \nn \\
&=& \f{1}{\left[ E^+_{\bk} - E^-_{\bk} - i\lk \g^+-\g^- \rk \right]^2} 
\f{E^+_{\bk}+E^-_{\bk}-2\mu -i\lk \g^+ + \g^- \rk}{\lk E^+_{\bk}-\mu -i \g^+ \rk \lk E^-_{\bk}-\mu -i \g^- \rk} \label{eq:IIa-1} \\ 
&+& \f{2}{\left[ E^+_{\bk} - E^-_{\bk} - i\lk \g^+-\g^- \rk \right]^3} \ln \f{E^+_{\bk} -\mu-i\g^+}{E^-_{\bk} -\mu-i\g^-}. \label{eq:IIb-1}
\end{\eq}
\end{widetext}

We analyze eqs. (\ref{eq:IIa-1}) and (\ref{eq:IIb-1}) when $\g$ is small:
Since ${\rm Im}[ \ln(x \pm i\gamma)] = \mp \pi \theta(x)$, the imaginary part of eqs. (\ref{eq:IIa-1}) and (\ref{eq:IIb-1}) is approximated as
\begin{\eq}
\text{Im} \left\{ \text{eq.} (\ref{eq:IIa-1}) \right\} &\approx& \f{\pi \delta\lk \mu-E^-_{\bk} \rk}{\lk E^+_{\bk} -E^-_{\bk} \rk^2}, \\
\text{Im} \left\{ \text{eq.} (\ref{eq:IIb-1}) \right\} &\approx& \f{2\pi \tht \lk \mu-E^-_{\bk}\rk}{\lk E^+_{\bk}-E^-_{\bk} \rk^3}, \label{eq:ImIIb}
\end{\eq}
%Note that $\tht\lk \mu-E^+_{\bk} \rk$ in eq. (\ref{eq:ImIIb}) vanishes since 
for Ce-compounds, where the Fermi level lies
under $E^f$, as shown in Fig. \ref{fig:band}.
Substituting above equations into eq. (\ref{eq:sxyII-2}), $\sxy^{IIa}$ and $\sxy^{IIb}$ is given by
\begin{\eq}
\sxy^{IIa}&=& \f{e}{2\pi N} \sum_{\bk} \f{52}{7} |V_f|^2 \f{1}{k} \f{\rd \eps_{\bk}}{\rd k} \f{\pi \delta\lk \mu-E^-_{\bk} \rk}{\lk E^+_{\bk}-E^-_{\bk} \rk^2}, \label{eq:IIa-2} \\
\sxy^{IIb}&=& \f{-e}{2\pi N} \sum_{\bk} \f{52}{7} |V_f|^2 \f{1}{k} \f{\rd \eps_{\bk}}{\rd k} \f{2\pi\tht\lk\mu-E^-_{\bk} \rk}{\lk E^+_{\bk}-E^-_{\bk} \rk^3}. \label{eq:IIb-2}
\end{\eq}
We will explain in Appendix \ref{App-A}
how to perform the $\bk$-summations in eq. (\ref{eq:IIb-2}).
In case of $|V_f|^2/(E^f-\mu) \gg 1$, final expressions for $\sxy^{IIa}$ and $\sxy^{IIb}$ are obtained as 
\begin{\eq}
%\sxy^{zIIa}&=& -\f{3|V_{k_F}|^2}{(E^+_{k_F}-\mu)^2} a_c^{-1} \sxy^{zI},  \\
\sxy^{IIa}&=& -\Lambda_{k_F} \sxy^{I}, \label{eq:IIa-va}  \\
\sxy^{IIb}&=& \sxy^{I}. \label{eq:IIb-f} \label{eq:IIb-va}
\end{\eq}
Here $\Lambda_{k_F}\equiv\f{|V_f|^2}{(E^+_{k_F}-\mu)^2} a_c^{-1}$, and $\displaystyle a_c^{-1}\equiv \left. \f{d \eps_{\bk}}{d E^-_{\bk}} \right|_{E^-_{\bk}=\mu}=1+\f{|V_f|^2}{(\mu-E^f)^2}$. 
Considering the relation $E^+_{k_F}\approx \eps_{k_F}$ in Fig. \ref{fig:band}, it is straight forward to show that 
$\Lambda_{k_F} \sim 1$ up to $O((\Delta^2/|V_f|)^2)$. 
In this case, we obtain the 
following relationships for small $\g$:
\begin{\eq}
\sxy^{I} &\sim& \sxy^{IIb} \sim -\sxy^{IIa}, \label{eq:I-II1} \\
\sxy^{I} &\gg& \sxy^{II}. \label{eq:I-II2}
\end{\eq}

Therefore, two Fermi sea terms $\sxy^{IIa}$ and $\sxy^{IIb}$ almost cancel,
and as a result, the Fermi surface term $\sxy^I$ gives a dominant contribution 
to the SHC \cite{Kontani06,Kontani-Ru,Kontani-Pt,Tanaka-4d5d}.
Note that the same relations also hold for the OHC, and the total 
OHC is mainly given by the Fermi surface term.

%%%%%%%%%%%%%%%%%%
% Discussions
%%%%%%%%%%%%%%%%%%
\section{\label{sec:4} Discussions}

\subsection{\label{subsec:4-1} SHC and OHC in Yb-compound system}

%SHC in Yb
Now, we discuss the SHC for $J=7/2$, which corresponds to the case in Yb-compound systems.

To perform $M,\sg$-summations, we use the following relations for $J=7/2$:
\begin{\eq}
&&\sum_{M \sg} M^2 \V2 =  \f{V_f}{2} \left( 1 + 30\sin^2\th \right), \label{eq:J72Ms1}  \\
&&\sum_{M \sg} \sg^2 \V2 = 2 V_f,  \\
&&\sum_{M \sg} M \sg \V2 = V_f \left( 1 + 3\sin^2\th \right). \label{eq:J72Ms2}
\end{\eq}

By using the above relationships, we can perform the calculation of $\sxy^{I}$
by following section \ref{subsec:3-1}. 
As a result, $\sxy^{I}$ for $J=7/2$ takes a large positive value as
\begin{\eq}
\sxy^{I}&=&e\f{15}{14}\f{k_{\rm F}}{2\pi^2} N_{FS}
 \nonumber \\
&=&\f{e}{2\pi a}\f{15}{14} N_{FS}.
 \label{eq:SH-Yb}
\end{\eq}
The second line in eq. (\ref{eq:SH-Yb}) is obtained 
by putting $k_{\rm F}=\pi/a$.
This result suggests that SHCs in Yb-compound heavy fermion systems
take similar large positive values.
We can also calculate the Fermi sea for $J=7/2$ 
by following section \ref{subsec:3-2}.
Then, we recognize the relationship in 
eqs. (\ref{eq:I-II1}) and (\ref{eq:I-II2}) for $J=7/2$.

%

%%%%%
%OHC
%%%%%
In the same way, the OHC for $J=7/2$ state is given by
\begin{\eq}
\sigma^{I}_{\rm OH} = \f{14}{5} \sxy^I.
 \label{eq:O-Yb}
\end{\eq} 

Therefore, we note that the sign of SHC is negative for $J=5/2$, it is positive for $J=7/2$, whereas the OHC is positive for both cases.
These facts are consistent with the results obtained in 4$d$- and 5$d$ transition
metals \cite{Tanaka-4d5d, Kontani-OHE}.
In section \ref{subsec:4-2}, we will show that the sign of SHC is equal to the sign of 
the spin-orbit polarization $\langle {\bm l} \cdot {\bm s} \rangle_{\mu}$\cite{Kontani-OHE}.

\subsection{\label{subsec:4-2} Orbital Aharonov-Bohm Phase Factor}

%%%%%%%%%%%%%%%%%%%%%%%%%%%%%%%%%%%%%%%%%%
\begin{figure}[!htb]
\includegraphics[width=.7\linewidth]{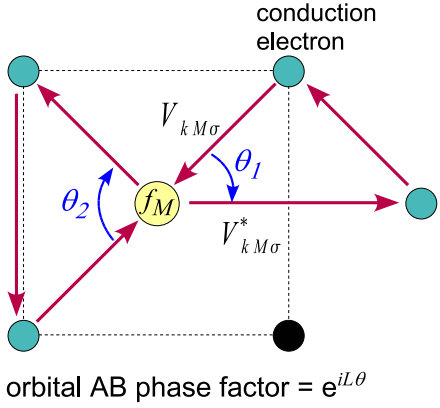}
\caption{\label{fig:ABphase} Effective Aharonov-Bohm phase in two-dimensional
OD-PAM. 
} 
\end{figure}
%%%%%%%%%%%%%%%%%%%%%%%%%%%%%%%%%%%%%%%%%%

In previous sections, we have discussed the SHE based on the OD-PAM
using the Green function method. 
In this section, we give an intuitive explanation for the origin of
the huge SHE in heavy fermion systems.
For this purpose, we consider the two orbital model with $M=\pm 5/2$,
assuming the strong crystalline field.
In the case of $J=5/2$, 
the $c$-$f$ mixing potential is given by
$V_{M \sg} (\hat \br) \propto \left\{ \sqrt{6} Y^{-3\sg}_{3} (\hat \br) 
\delta_{M,-5/2\sg} + Y^{2\sg}_{3} (\hat \br) \delta_{M,5/2\sg} \right\}$
in the real space representation:
If we drop the second term, it can be approximated as 
$V_{M \sigma} (\br) \propto Y_3^{-3 \sigma} (\hat \br) \propto {\rm e}^{-3i \sigma \phi_r}$, where $\phi_r = \tan^{-1} (y/x)$.

In Fig. \ref{fig:ABphase},  
two examples of the clockwise motion of the conduction electron
along the nearest three sites [$f_M \rightarrow c \rightarrow c \rightarrow 
f_M$] are shown.
Here, $\theta_i$ represents the angle between the incoming and outgoing electron.
Therein, the electron acquires the phase factor ${\rm e}^{-3 \sg \theta}$ 
due to the angular dependence of the $c$-$f$ mixing potential in real space,
$V_{M \sigma} (\hat \br)$. 
This phase factor can be interpreted as the ``orbital AB phase factor" 
at the $f$-site, which works as
the effective magnetic flux $(-3\sg\theta/2\pi) \phi_0$ through the
area of the triangle. Here, $\phi_0 = 2\pi \hbar/e$ is the flux quantum.
On the other hand, $V_{M \sg}$ 
is approximately given by
$V_{M \sg}(\hat \br) \propto {\rm e}^{3i\sg \phi_r}$ for $J=7/2$.
In this case, the effective magnetic flux per triangle is
$(3\sg\theta/2\pi) \phi_0$,
which is opposite to that for $J=5/2$.

In summary, a conduction electron acquires the spin-dependent 
``orbital AB phase factor", which originates from
the spin-dependent $c$-$f$ hybridization in the presence of strong SOI.
This is the origin of the huge SHE in heavy fermion systems.
This consideration also explains the sign difference of the SHC 
between Ce- and Yb-compounds.
Thus, the origin of the SHE in heavy fermion systems 
is well understood based on the simplified two-orbital model.

\subsection{\label{subsec:4-3} The relationship between the intrinsic and side-jump terms }

So far, we have studied the OD-PAM with translational invariance, 
and found that huge intrinsic SHC emerges.
Here, we consider the depletion of $f$-electron.
The quasiparticle damping rate $\gamma$ increases in proportion
to the depletion ratio $x$.
In the case of $x \ll 1$, the intrinsic SHC is independent of $x$
if $\gamma$ is smaller than the band splitting \cite{Kontani-Pt,Tanaka-4d5d}.
In addition to the intrinsic term, the depletion may induce the extrinsic terms,
that is, skew scattering term $\sxy^{\rm skew}$ and side-jump term $\sxy^{\rm sj}$.

In the dilute limit where $1-x \ll 1$, intrinsic term does not exist.
In this case, present authors had studied 
the extrinsic SHE based on the orbitally 
degenerate single-impurity Anderson model \cite{Tanaka-NJP}.
For $k_{F}= \pi/a$ ($a$ is a lattice spacing),
the expressions for skew scattering and side-jump terms
are obtained as
\begin{\eq}
\sxy^{\rm sj}= \f{e}{2\pi a}\f{2}{3} \ls, 
 \label{eq:Ssj}\\
\sxy^{\rm skew}= \f{e}{2\pi a} \delta_2 \f{1}{\gamma}\ls,
\end{\eq}
for both $J=5/2$ ($\ls=-2$) and $J=7/2$ ($\ls=3/2$).
Here, $\delta_2$ is a phase shift for $d$ partial wave.
From the above equation, we find that the extrinsic term is proportional to the spin-orbit polarization $\ls$.
In these two Anderson models, 
both intrinsic term $\sxy^{\rm int}$ and side-jump term $\sxy^{\rm sj}$ originate from the anomalous velocity that arises from the $\bk$-derivative of the phase factor in the mixing potential.
Here, we compare eqs. (\ref{eq:I-va}), (\ref{eq:SH-Yb}), and ({\ref{eq:Ssj}).
Very interestingly, the following relationship holds in a accuracy of 
$\pm7.2$\%:
\begin{\eq}
\sxy^{\rm int} \approx \sxy^{\rm sj} \label{eq:intsj}
\end{\eq}
This fact indicates unexpected close relationship between 
the intrinsic term and the extrinsic side-jump term, and therefore
it would be very difficult to distinguish these two mechanisms experimentally.
This fact would be the reason why
intrinsic (or side-jump) term are widely observed from single crystals to 
polycrystal or amorphous compounds.

\section{\label{sec:5} Summary}

In this paper, we studied the intrinsic SHE and OHE based on the OD-PAM.
We derived the analytical expression for the intrinsic SHC and OHC based on 
the linear response theory.
Both SHC and OHC are mainly given by the Fermi surface term ($I$). 
The obtained results for Ce-compounds ($J=5/2$) are given by 
eqs. (\ref{eq:I-va}) and (\ref{eq:O-Ce}),
and those for Yb-compounds ($J=7/2$) are given by 
eqs. (\ref{eq:SH-Yb}) and (\ref{eq:O-Yb}).
The SHCs for both compounds are approximately 
expressed by eq. (\ref{eq:Ssj}).
These results suggests that SHCs in $Ce$- ($Yb$-) compound 
heavy fermion systems take similar large negative (positive) values;
$2000\sim 3000 \hbar e^{-1} \Omega^{-1}$cm$^{-1}$ in magnitude.
The mechanism of the huge SHE and OHE in $f$-electron systems is the
``orbital AB effect'', which is given by the spin-dependent 
Berry phase induced by the complex $f$-orbital wavefunction.
%orbital AB effect as in $d$-electron systems. 
Therein, the SHC is proportional to the spin orbit polarization $\ls$.
The SHC in $f$-electron systems will be measurable 
by using recently developed fabrication technique 
of high quality heavy fermion thin film \cite{Shishido}.

Here, we briefly comment on the effect of the Coulomb interaction $U$.
In the present study, we have calculated the SHC with $U=0$.
In the PAM, the effect of the self-energy correction 
is represented by the renormalization of the mixing potential 
$V_{\bk M \sg}\rightarrow\sqrt{z} V_{\bk M \sg}$, where $z^{-1} \equiv
 1-\f{\rd}{\rd \eps} \Sigma(\eps) =m^{\ast}/m $ is the 
renormalization factor due to the self-energy
 \cite{Gutzwiller,Rice-Ueda}.
Since the SHC obtained in this study is independent of $V_{\bk M \sg}$, 
the SHC will be independent of the mass-enhancement 
due to Coulomb interaction.
(In contrast, the AHE under the magnetic field
is proportional to the magnetic susceptibility $\chi^S \propto m^*/m$.)
Next, we discuss the CVC due to Coulomb interaction.
In ref. \cite{Kontani94}, it was proved that 
the CVC by $U$ does not give rise to the skew scattering term, 
and thus its quantitative effect on the SHE
is expected to be small \cite{Kontani94}.
%For the same reason, the CVC is expected to be unimportant for the SHE. 
%As is the case with the SHE, the vertex correction is negligible small in
%quantitative discussion.
However, the CVC due to spin fluctuations might be significant
in nearly quantum-critical-point \cite{Kontani-review}.
This is an important future issue.

%%%%%%%%%%%%%%%%%%
% Acknowledgment
%%%%%%%%%%%%%%%%%%
\begin{acknowledgements}

The authors are grateful to D. S. Hirashima, J. Inoue, 
T. Terashima, Y. Matsuda, Y. Otani, T. Kimura, and K. Yamada 
for fruitful discussions.
This work has been supported by a Grant-in-Aid for Scientific Research
on Innovative Areas gHeavy Electronsh (No. 20102008) of
The Ministry of Education, Culture, Sports, Science, and Technology, Japan.

\end{acknowledgements}

%%%%%%%%%%%%%%%%%%%%
% references
%%%%%%%%%%%%%%%%%%%%

\appendix

%%%%%%%%%%%%%%%%%%%%
% Appendix A
%%%%%%%%%%%%%%%%%%%%
\section{\label{App-A} Derivation of eq. (\ref{eq:IIb-f}) }

Here, we explain the way we performed the $\bk$-summations in eq. (\ref{eq:IIb-2}), and derive eq. (\ref{eq:IIb-f}).
In performing the $\bk$-summations analytically, we 
assumed that the density of state $N(\w)$ for conduction electron is constant:
$\sum_{\bk}=N(0)\int d\eps_{\bk}$.
Then, 
\begin{\eq}
&&\int^{X}_{-\infty}\f{d\eps_{\bk}}{ \left( E^+_{\bk}-E^-_{\bk} \right)^3} \nn \\
&=& \int^{X}_{-\infty} \f{d\eps_{\bk}}{\left[ (\eps_{\bk}-E^f)^2 +4|V_{f}|^2 \right]^{3/2} } \nn \\
&=& \f{1}{4|V_{f}|^2} \left\{ \f{\tilde X}{\tilde X^2 + 4|V_{f}|^2} + 1 \right\}, \label{eq:IIb-int}
\end{\eq}
where $\tilde X\equiv \mu- E^f -\f{|V_{f}|^2}{\mu-E^f}$.
When $|V_{f}|^2/(E^f-\mu) \gg 1$, the first term in the bracket in eq. (\ref{eq:IIb-int}) is approximated as $\approx 1$.  
As a result, $\sxy^{IIb}$ is given by
\begin{\eq}
\sxy^{IIb}&=&-\f{e}{2\pi a} \f{26}{21} N_{FS}=\sxy^{I}.
\end{\eq}

%%%%%%%%%%%%%%%%%%%%
% Appendix B
%%%%%%%%%%%%%%%%%%%%
\section{\label{App-JJ} Calculations of the term proportional to 
$\rd_{x}V_{\bk}\rd_y V_{\bk}$.}
% for both $J^S_{x}$ and $J^C_{y}$are composed of the anomalous velocity }

%\subsection{\label{subsec:3-3} SHC for both $J^S_{x}$ and $J^C_{y}$ are 
%composed \\ of the anomalous velocity}

%%%%%%%%%%%%%%%%%%%%%%%%%%%%%%%%%%%%%%%%%%
\begin{figure}[!htb]
\includegraphics[width=.95\linewidth]{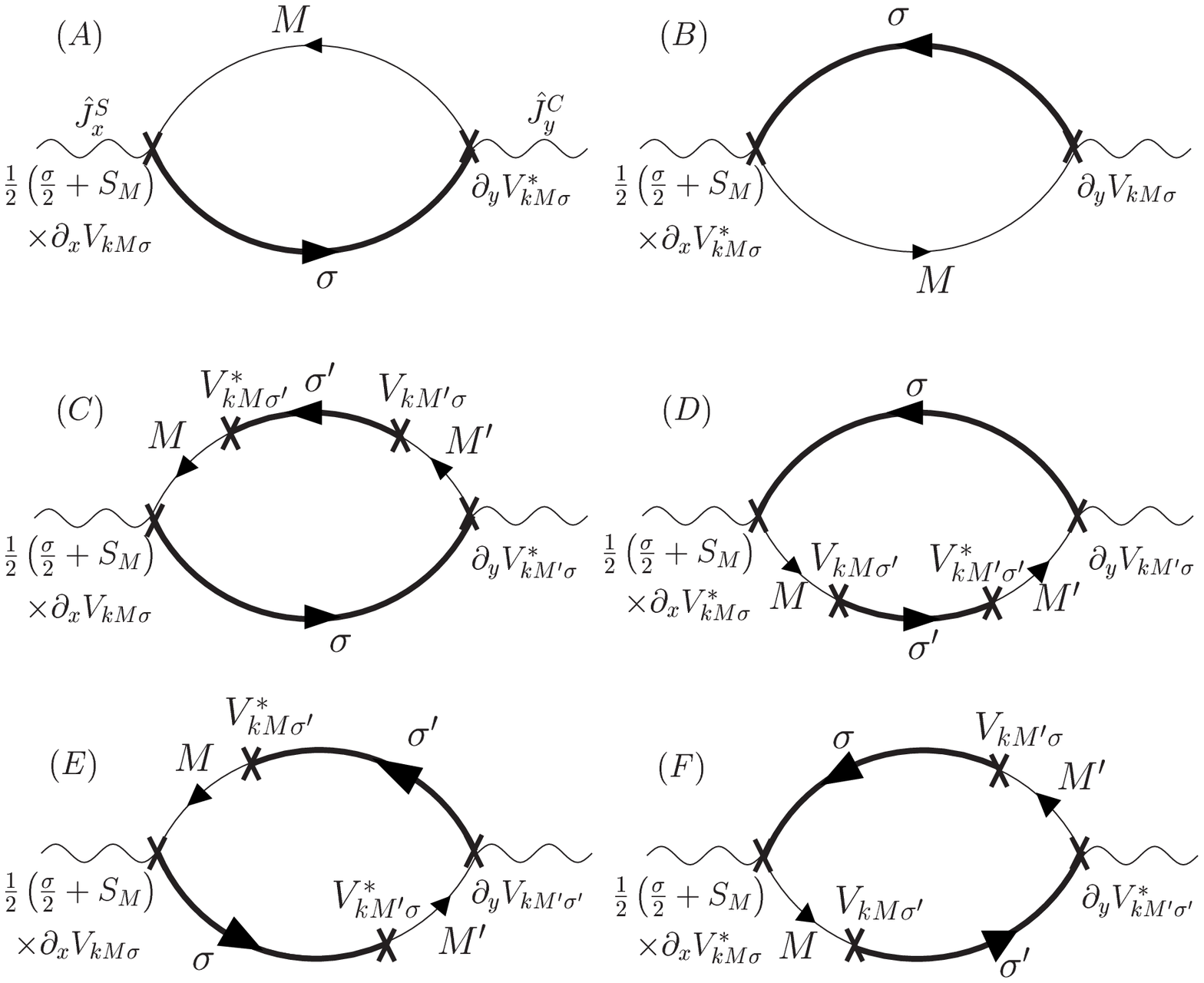} 
\caption{\label{fig:diagram2} The diagrammatic expression for the term
proportional to $\rd_x V_{\bk M \sg} \rd_y V_{\bk M \sg}$.
%for $\hat J^S_x$ and $\hat J^C_y$ are composed of the anomalous velocity.
%(b) The diagrams which cancels each other.
} 
\end{figure}
%%%%%%%%%%%%%%%%%%%%%%%%%%%%%%%%%%%%%%%%%%

%M,\sigma-summations
In the main text, we have calculated the term proportional to $\rd_{\mu} V \rd_{\nu}\eps_{\bk}$, and explained that it gives a dominant contribution to the SHC.
In this appendix, we derive the SHC given by $\rd_{x}V_{\bk}\rd_y V_{\bk}$,
and show that it is very small and negligible.
%for both spin and charge currents are composed of anomalous velocity.
In this case, to perform the $M,\sg$-summations, we use the following relations:
\begin{\eq}
&&\sum_{M,\sg} \f{1}{2} \left( \f{\sg}{2} - \f{M}{7} \right) \f{\rd \Vs}{\rd k_x}\f{\rd \Vs^{\ast}}{\rd k_y} = i \f{12}{7} \cos^2 \th \f{1}{k^2} |V_f|^2, \nn \\
&&\sum_{M,M',\sg,\sg'}\f{1}{2} \left( \f{\sg}{2} - \f{M}{7} \right) \f{\rd \Vs}{\rd k_x} V^{\ast}_{\bk M\sg'} V_{\bk M' \sg'} \f{\rd V^{\ast}_{\bk M'\sg}}{\rd k_y} \nn \\
&&= i \f{16}{7} \cos^2\th \f{1}{k^2}  |V_f|^4 , \nn \\
&&\sum_{M,M',\sg,\sg'}\f{1}{2} \left( \f{\sg}{2} - \f{M}{7} \right) \f{\rd \Vs}{\rd k_x} V^{\ast}_{\bk M\sg'} V^{\ast}_{\bk M' \sg} \f{\rd V_{\bk M'\sg'}}{\rd k_y} \nn \\
&&= -i \f{16}{7} \cos^2\th \f{1}{k^2} |V_f|^4.  \label{eq:sum-Ms} 
\end{\eq}

Here, we first perform the calculation for the Fermi surface term.
By using the above relationship shown in eqs. (\ref{eq:sum-Ms}),
the SHC given by (A)-(D) in Fig. \ref{fig:diagram2} is given by
\begin{\eq}
\sxy^{I (A-D)} &=& \f{e}{2\pi} \f{8}{7} \sum_{\bk} \f{1}{k^2} |V_f|^2 \text{Im}  \left\{ G^{fR}(0)G^{cA}(0) \right\} \label{eq:I-AB} \\
&+& \f{e}{2\pi} \f{64}{21} \sum_{\bk} \f{1}{k^2}  |V_f|^4  \text{Re} G^{fR}(0) \text{Im}G^{fR}(0) |G^c(0)|^2. \label{eq:I-CE} \nn \\
\end{\eq}
The diagrammatic expressions for eqs. (\ref{eq:I-AB}) and (\ref{eq:I-CE}) 
are respectively given by 
(A) and (B), and (C) and (D) in Fig. \ref{fig:diagram2}.
The contributions from the diagrams (E) and (F) turn out to cancel out.

When $\g$ is small, we obtain a following relationship: 
\begin{\eq}
&&\text{Im} \left\{ G^{fR}(0) G^{cA}(0) \right\} \approx \f{\pi \delta(\mu-E_{\bk})}{\mu-E^f}, \label{eq:GfGc} \\
&&\text{Re} G^{fR}(0) \text{Im}G^{fR}(0) |G^c(0)|^2 \approx -\f{1}{|V_f|^2}\f{\pi \delta(\mu-E_{\bk})}{\mu-E^f}. \nn \\
\end{\eq}
Substituting the above equations into eqs. (\ref{eq:I-AB}) and (\ref{eq:I-CE}), and performing the $\bk$-summation, we obtain 
\begin{\eq}
\sxy^{I} %&=& -\f{e}{2 \pi a} \f{2}{7}\alpha + \f{e}{2\pi a} \f{16}{21} \alpha \nn \\
&=& \f{e}{2\pi a} \f{10}{21} \alpha N_{FS}, \label{eq:FSVV}
\end{\eq}
where $\alpha$ is defined by $|\mu -\eps_{k_F}|=\alpha \eps_{k_F}$.

In a similar way, we calculate $\sxy^{IIa}$ and $\sxy^{IIb}$.
After performing the $M,\sigma$-summations using eq. (\ref{eq:sum-Ms}), we obtain the 
following expression of the Fermi sea term for (A)-(D) in Fig. \ref{fig:diagram2}:
\begin{widetext}
%diagrams A+B
\begin{\eq}
\sxy^{II (A-D)} &&=-\f{e}{4\pi} \f{48}{21} \sum_{\bk} \f{1}{k^2} |V_f|^2 \text{Im} \left\{ \int^{0}_{-\infty} d\w \f{\rd G^{fR}(\w)}{\rd \w} \cdot G^{cR}(\w) - G^{fR}(\w)\f{\rd G^{cR}(\w) }{\rd \w}  \right\}  \\
&&-\f{e}{2\pi a} \f{128}{21} \sum_{\bk} \f{1}{k^2} |V_f|^4 \text{Im} \left\{ \int^{0}_{-\infty} d\w \f{\rd G^{fR}(\w)}{\rd \w} \cdot G^{fR}(\w) \cdot \left(G^{cR}(\w) \right)^2 \right\}.
\end{\eq}
\end{widetext}
As explained in section \ref{subsec:3-2}, after performing $\w$-integration, 
above expressions are rewritten as follows for small $\gamma$:
\begin{widetext}
\begin{\eq}
\sxy^{IIa (A-D)} &=& \f{e}{2\pi} \f{24 \pi}{21} \sum_{\bk} \f{1}{k^2} \f{|V_{f}|^2}{E^+_{\bk}-E^-_{\bk}} \delta(\mu - E^-_{\bk}) - \f{e}{2\pi} \f{64 \pi}{21} \sum_{\bk} \f{1}{k^2} 
\f{ |V_{f}|^4 }{(E^f - E^-_{\bk})(E^+_{\bk} -E^-_{\bk})^2} \delta(\mu-E^-_{\bk}), \nn \\ \label{eq:IIa-bk}
\end{\eq}
\begin{\eq}
\sxy^{IIb (A-D)} &=& -\f{e}{2\pi} \f{48\pi}{21} \sum_{\bk} \f{1}{k^2}  \f{|V_{f}|^2}{(E^+_{\bk}- E^-_{\bk})(E^f - E^-_{\bk})} \theta(\mu - E^-_{\bk}) \nn \\
&& + \f{e}{2\pi } \f{64 \pi}{21} \sum_{\bk} \f{1}{k^2} |V_{f}|^4 \theta(\mu-E^-_{\bk})  \left[ \f{1}{(E^f-E^-_{\bk})(E^+_{\bk}-E^-_{\bk})^2} + \f{2}{(E^f-E^-_{\bk})(E^+_{\bk}-E^-_{\bk})^3} \right]. \nn \\ \label{eq:IIb-bk}
\end{\eq}
\end{widetext}
Performing the $\bk$-summations in eq. (\ref{eq:IIa-bk}), 
and as a result, we obtain the following expressions for $\sxy^{IIa}$:
\begin{\eq}
%\sxy^{zIIa} &=& \f{e}{2\pi a}  \f{3|V_{\bk}|^2}{(E^+_{\bk}-E^-_{\bk})^2} a^{-1}_c \left[
%\f{2}{7} \beta - \f{16}{21} \alpha \right]
\sxy^{IIa (A-D)} &=& \f{e}{2\pi a}  \Lambda_{k_F} \left[
\f{2}{7} \beta - \f{16}{21} \alpha \right] N_{FS}, \label{eq:IIaVV}
\end{\eq}
where $\beta$ is defined by $|E^{+}_{k_F}-\mu|=\beta\eps_{k_F}$.
As recognized in Fig. \ref{fig:band}, the relation
$\alpha\approx\beta\sim \f{1}{2}$ is satisfied 
since $E^+_{k_F} \approx \eps_{k_F}$ is satisfied in the present model. 
Since the relation $\Lambda_{k_F}=1+O((\Delta/V_f)^2)$ holds well  
as discussed in ref. \ref{subsec:3-2},
$\sxy^{IIa}$ is given by
\begin{\eq}
\sxy^{IIa (A-D)} &=& -\f{e}{2\pi a} \f{10}{21}N_{FS} \alpha.
\end{\eq} 

To perform $\bk$-summations in eq. (\ref{eq:IIb-bk}), we use the following approximation:
$\displaystyle \sum_{\bk} \f{1}{k^2} \approx N(0)\f{1}{k_F^2}\int d\eps_{\bk}$.
Then, $\sxy^{IIb}$ is given by
\begin{\eq}
\sxy^{IIb} &=& -\f{e}{2\pi a} \left[ \left\{ -\f{1}{14} + \f{1}{21} \right\} \alpha + O \left( \f{\Delta}{|V_{k_F}|} \right) \right] N_{FS} \nn \\
&=&-\f{e}{2\pi a} \f{1}{42} N_{FS} \alpha. \label{eq:IIbVV}
\end{\eq}

Finally, we obtain the final expressions for 
$\sxy^{I}, \sxy^{IIa}$, and $\sxy^{IIb}$ are given by the summations of eqs. (\ref{eq:I-va}) and (\ref{eq:FSVV}), eqs. (\ref{eq:IIa-va}) and (\ref{eq:IIaVV}), and eqs. (\ref{eq:IIb-va}) and (\ref{eq:IIbVV}), respectively.
\begin{\eq}
\sxy^{I-\rm{tot}} &=& -\f{e}{2\pi a} \left( \f{26}{21} -\f{10}{21}\alpha \right) N_{FS}, \label{eq:I} \\
\sxy^{IIa-\rm{tot}} &=& \f{e}{2\pi a} \left( \f{26}{21} -\f{10}{21}\alpha \right) N_{FS}, \label{eq:IIa} \\
\sxy^{IIb-\rm{tot}} &=& -\f{e}{2\pi a} \left( \f{26}{21} + \f{1}{42}\alpha \right) N_{FS}. \label{eq:IIb}
\end{\eq}
Here in eqs. (\ref{eq:I}) -(\ref{eq:IIb}), the terms that is proportional to $\alpha$ is
given in Fig. \ref{fig:diagram2}.
In total, the SHC is given as
\begin{\eq}
\sxy^{\rm{tot}}=-\frac{e}{2 \pi a} \left( \frac{26}{21} + \frac{1}{42}\alpha \right) N_{\rm{FS}}.
\label{eq:sxy-tot}
\end{\eq} 
In eq. (\ref{eq:sxy-tot}), the factor ${26}/{21}$ and 
$\alpha/42$ in the bracket come from the terms with
$\rd_{\mu} V_{\bk}\rd_{\nu} \eps_{\bk}$ and the terms with
$\rd_{\mu} V_{\bk} \rd_{\nu} V_{\bk}$, respectively.
Since $\alpha\sim1/2$, 
the terms proportional to $\rd_{\mu} V_{\bk}\rd_{\nu} \eps_{\bk}$ 
shown in Fig. \ref{fig:diagram} gives a dominant contribution.

%%%%%%%%%%%%%%%%%%%%%%%%%%%%%%%%%%%%%%%%%

\end{document}